\documentclass[12pt,oneside,a4paper]{article}%
\usepackage{amsfonts}
\usepackage{amssymb}
\usepackage{amscd}
\usepackage{amsmath}
\usepackage{float}
\usepackage{tikz}
\usepackage{amsthm}
\usepackage{xkeyval}
\usepackage[top=2cm,bottom=2cm,right=2cm,left=3cm]{geometry}
\usepackage{graphicx}%
\providecommand{\U}[1]{\protect\rule{.1in}{.1in}}
\newlength{\bibitemsep}
\newlength{\bibparskip}
\let\oldthebibliography\thebibliography
\renewcommand\thebibliography[1]{  \oldthebibliography{#1}  \setlength{\parskip}{\bibitemsep}  \setlength{\itemsep}{\bibparskip}}
\providecommand{\U}[1]{\protect\rule{.1in}{.1in}}
\usetikzlibrary{calc}
\newtheorem{theorem}{Theorem}
\newtheorem{lemma}[theorem]{Lemma}
\newtheorem{corollary}[theorem]{Corollary}

\newtheorem*{proposition*}{Proposition}
{\theoremstyle{definition}

\newtheorem{example}[theorem]{Example}

\newtheorem*{notation*}{Notation}
}
\begin{document}

\noindent{\LARGE \textbf{Transforming St\"ackel Hamiltonians of Benenti type
to polynomial form }}

\vspace{5mm}

\noindent{\large Jean de Dieu Maniraguha$^{\dag}$, Krzysztof Marciniak$^{\ddag
}$ and C\'elestin Kurujyibwami$^{\dag}$ }

\vspace{2mm}

\vspace{2mm}

\noindent\textit{$^{\dag}$\,College of Science and Technology,\, University of
Rwanda,\, P.O.\,Box: 3900, Kigali, Rwanda}\newline$\phantom{^\dag}$\,E-mail:
jadomanir99@gmail.com\newline$\phantom{^\dag}$\,E-mail: celeku@yahoo.fr

\vspace{2mm}

\noindent\textit{$^{\ddag}$\, Department of Science and Technology,\, Campus
Norrk\"oping,\,Link\"oping University,\newline$\phantom{^\ddag}$\, 601-74
Norrk\"oping, Sweden,}\newline$\phantom{^\ddag}$\, E-mail: krzma@itn.liu.se

\begin{center}
January 2, 2021
\end{center}

\begin{abstract}
In this paper we discuss two canonical transformations that turn St\"{a}ckel
separable Hamiltonians of Benenti type into polynomial form: transformation to
Vi\`{e}te coordinates and transformation to Newton coordinates. Transformation
to Newton coordinates has been applied to these systems only very recently and
in this paper we present a new proof that this transformation indeed leads to
polynomial form of St\"{a}ckel Hamiltonians of Benenti type. Moreover we
present all geometric ingredients of these Hamiltonians in both Vi\`{e}te and
Newton coordinates.

\end{abstract}


\pagenumbering{arabic}

\section{Introduction}

The aim of this paper is to investigate two canonical transformations of the
phase space to coordinates in which the so called St\"{a}ckel separable
systems of Benenti type attain a polynomial form, as well as to present all
geometric objects, related with such systems (the pseudo-Riemannian metric
tensor and its Killing tensors as well as the conformal Killing tensor,
present in the Hamiltonians of the system) in these new coordinates.

St\"{a}ckel systems constitute an important family of quadratic in momenta
Hamiltonian systems that are separable, in the sense of Hamilton-Jacobi
theory, in orthogonal coordinates. These systems were introduced by Paul
St\"{a}ckel in~\cite{Stackel1891}, where he presented the conditions for
separability of Hamilton-Jacobi equation of a natural Hamiltonian system (that
is a system of the form $H=K+V$ where $K$ is a quadratic in momenta form and
$V$ is a potential defined on the underlying configurational space of the
system) in orthogonal coordinates, see for example~\cite{waksjo2003} for a
comprehensive review of this subject. St\"{a}ckel systems can most
conveniently be obtained from the separation relations~\cite{sklyanin1995}
that are linear in the Hamiltonians $H_{i}$ and quadratic in momenta $\mu_{i}%
$. Further specification of ingredients in these separation relations lead to
so called Benenti systems (see the next section for all the necessary
definitions and details).

The obtained St\"{a}ckel (or Benenti) Hamiltonians $H_{j}$, as well as their
geometric components, are usually given by complicated rational functions, if
written in the canonical coordinates in which they were originally created
through separation relations. In literature, two maps turning Benenti systems
into polynomial form are known: the map to the so called Vi\`{e}te coordinates
\cite{blaszak2019} and the map to the so called Newton
coordinates~\cite{buchstaber2018}, the second map discovered only recently. In
this paper we improve the results obtained in~\cite{buchstaber2018} by
presenting an alternative, much simpler, proof of its main result using the
direct map between Vi\`{e}te coordinates and Newton coordinates. We also
present the explicit form of all the geometric structures that are present in
the Benenti Hamiltonians in Newton coordinates. These results are new.

\section{St\"ackel systems}

\label{Sec.Stackel sytems} Consider a $2n$-dimensional manifold $\mathcal{M}$
equipped with a Poisson bracket $\pi$. Suppose also that $(\lambda
,\mu)=(\lambda_{1},\ldots,\lambda_{n},\mu_{1},\ldots,\mu_{n})$ are global
Darboux coordinates on $\mathcal{M}$ (i.e, $\left\{  \lambda_{i},\lambda
_{j}\right\}  =\left\{  \mu_{i},\mu_{j}\right\}  =0$ for all $i,j=1,\ldots,n$
while $\left\{  \lambda_{i},\mu_{j}\right\}  =\delta_{ij}$). A set of
algebraic equations of the form
\begin{equation}
\varphi_{i}(\lambda_{i},\mu_{i},a_{1},\ldots,a_{n})=0\text{,\quad}i=1,\ldots,n
\label{0.4}%
\end{equation}
is called separation relations if it is globally solvable (except possibly for
a union of lower dimensional submanifolds) with respect to the parameters
$a_{j}\in\mathbb{R}$.

Among all possible separations relations (\ref{0.4}), a natural subclass
consists of the separation relations that are linear in the Hamiltonians
$H_{k}$:
\begin{equation}
\sum_{k=1}^{n}S_{ik}(\lambda_{i},\mu_{i})H_{k}=\psi_{i}(\lambda_{i},\mu
_{i}),\quad i=1,\dots,n. \label{S1}%
\end{equation}
Here $S_{ik}$ and $\psi_{i}$ are arbitrary smooth functions of two arguments
$(\lambda_{i},\mu_{i})$. The relations (\ref{S1}) are called the
\emph{generalized St\"{a}ckel separation relations} and the related dynamical
systems, obtained by solving (\ref{S1}) with respect to $H_{k}$, are called
the \emph{generalized St\"{a}ckel systems}. The matrix $S=\left[
S_{ik}(\lambda_{i},\mu_{i})\right]  $ is called a \emph{generalized
St\"{a}ckel matrix}. Although the restriction to separation relations linear
in $H_{k}$ seems to be very strong, it appears that an overwhelming majority
of all separable systems considered in the literature falls into various
subclasses of this class. The most important class of systems in (\ref{S1}) is
the class of \emph{classical St\"{a}ckel systems}, that is systems with the
matrix $S$ being a St\"{a}ckel matrix (so that \thinspace$S_{ik}%
=S_{ik}(\lambda_{i})$) and with $\psi_{i}$ being \emph{quadratic} in momenta
$\mu$:%

\begin{gather*}
S_{ik}(\lambda_{i},\mu_{i})=S_{ik}(\lambda_{i}), \quad\psi_{i}(\lambda_{i}%
,\mu_{i})=\frac12 f_{i}(\lambda_{i})\mu_{i}^{2}-\varphi_{i} (\lambda_{i}),
\end{gather*}
so that the separation relations (\ref{S1}) attain the form
\begin{equation}
\label{S45}\varphi_{i}(\lambda_{i})+\sum_{k=1}^{n} S_{ik}(\lambda_{i}%
)H_{k}=\tfrac{1} {2}f_{i}(\lambda_{i})\mu_{i}^{2},\quad i=1,\dots,n.
\end{equation}
The relations (\ref{S45}) are called \emph{St\"ackel separation relations}. A
particular St\"ackel system is thus defined by a choice of the St\"ackel
matrix $S_{ik}(\lambda_{i})$ and by a choice of $2n$ functions $f_{i}$ and
$\varphi_{i}$. Solving the relations (\ref{S45}) with respect to $H_{k}$ we
obtain $n$ quadratic in momenta functions (Hamiltonians) on $\mathcal{M}$
\begin{equation}
H_{r}=\frac12 \mu^{T}A_{r}\mu+V_{r} (\lambda)\text{, \quad}r=1,\ldots,n,
\label{1a}%
\end{equation}
where $A_{r}$ are $n\times n$ matrices given by
\[
A_{r}=\text{diag}\left(  f_{1}\left(  \lambda_{1}\right)  \left(
S^{-1}\right)  _{r1},\ldots,f_{n}\left(  \lambda_{n}\right)  \left(
S^{-1}\right)  _{rn}\right)  \text{, \quad}r=1,\ldots,n.
\]
As the Hamiltonians (\ref{1a}) are defined through separation relations, they
are in involution with respect to the canonical Poisson bracket on
$\mathcal{M}$.

There is a natural geometric interpretation of St\"{a}ckel systems given by
(\ref{1a}). If we factorize $A_{r}$ as $A_{r}=K_{r}G$, where
\[
G=A_{1}=\text{diag}\left(  f_{1}\left(  \lambda_{1}\right)  \left(
S^{-1}\right)  _{11},\ldots,f_{n}\left(  \lambda_{n}\right)  \left(
S^{-1}\right)  _{1n}\right)
\]
and
\[
K_{r}=\text{diag}\left(  \frac{\left(  S^{-1}\right)  _{r1}}{\left(
S^{-1}\right)  _{11}},\ldots,\frac{\left(  S^{-1}\right)  _{rn}}{\left(
S^{-1}\right)  _{1n}}\right)  \text{, \quad}r=1,\ldots,n
\]
(so that $K_{1}=I$) then we can interpret the matrix $G$ as a contravariant
form of a metric tensor on a manifold $\mathcal{Q}$ such that $\mathcal{M}%
=T^{\ast}\mathcal{Q}$ is the cotangent bundle to $\mathcal{Q}$. The
corresponding covariant metric tensor will be denoted by $g$ so that $gG=I$.
It can be shown that the matrices $K_{r}$ are then $(1,1)$-Killing tensors of
the metric $G$. For a fixed St\"{a}ckel matrix $S$ we have thus the whole
family of metrics $G$ parametrized by $n$ arbitrary functions $f_{i}$ of one
variable $\lambda_{i}$. The tensors $K_{r}$ are then Killing tensors for any
metric from this family. Thus, the St\"{a}ckel Hamiltonians $H_{r}$ in
(\ref{1a}) are geodesic Hamiltonians of a Liouville integrable system in the
Riemannian space $(\mathcal{M},g)$. Further, due to the linearity of the
separation relations (\ref{S45}), the functions $V_{r}(\lambda)$ on
$\mathcal{Q}$ are defined by the following separation relations
\[
\sum_{k=1}^{n}S_{ik}(\lambda_{i})V_{k}=-\varphi_{i}(\lambda_{i}),\text{ \quad
}i=1,\dots,n,
\]
and are called in literature separable potentials on $\mathcal{Q}$.

\section{St\"ackel systems of Benenti type}

\label{Sect.Stackel systems of Benenti type}

From now on we restrict ourselves to the case the St\"{a}ckel matrix $S$ in
(\ref{S45}) is of the very particular form $S_{ij}=\lambda_{i}^{n-j}$ or
explicitly:
\begin{equation}
S=\left(
\begin{array}
[c]{cccc}%
\lambda_{1}^{n-1} & \lambda_{1}^{n-2} & \ldots & 1\\
\vdots & \vdots & \vdots & \vdots\\
\lambda_{n}^{n-1} & \lambda_{n}^{n-2} & \ldots & 1
\end{array}
\right)  \label{SBen}%
\end{equation}
thus being a Vandermonde matrix. The corresponding St\"{a}ckel systems are
thus defined by separation relations of the form
\begin{equation}
\varphi_{i}(\lambda)+{\displaystyle\sum\limits_{j=1}^{n}}\lambda_{i}%
^{n-j}H_{j}=\frac{1}{2}f_{i}(\lambda_{i})\mu_{i}^{2}\quad i=1,\ldots,n,
\label{Ben}%
\end{equation}
and are called in literature \emph{Benenti systems}. Benenti systems have been
studied much in literature recently, see for example~\cite{blaszak2006,
blaszak2020} and references therein.

The inverse of $S$ as given by (\ref{SBen}) is given by the following lemma.

\begin{lemma}
\label{VDM}If $S$ is the $n\times n$ Vandermonde matrix given by
$S_{ij}=\lambda_{i}^{n-j}$ then
\[
\left[  S^{-1}\right]  _{ij}=-\frac{1}{\Delta_{j}}\frac{\partial\rho_{i}
}{\partial\lambda_{j}},
\]
where
\[
\rho_{i}=(-1)^{i}\sigma_{i}(\lambda),\quad\Delta_{j}= {\textstyle\prod
\limits_{k\neq j}} (\lambda_{j}-\lambda_{k})
\]
and where $\sigma_{r}\left(  \lambda\right)  $ are elementary symmetric polynomials.
\end{lemma}

By definition
\[
\sigma_{i}(\lambda)=\displaystyle\sum\limits_{1\leq j_{1}<\ldots<j_{i}\leq
n}\lambda_{j_{1}}\ldots\lambda_{j_{i}},\text{ \ }i=1,\ldots,n,
\]
so that
\[
\sigma_{0}=1\text{, \ }\sigma_{1}=\sum\limits_{i=1}^{n}\lambda_{i}\text{,
\ }\sigma_{2}=\sum\limits_{1\leq i<j\leq n}^{n}\lambda_{i}\lambda_{j}\text{,
}\ldots\text{, \ }\sigma_{n}={\textstyle\prod\limits_{i=1}^{n}}\lambda_{i}.
\]
Lemma \ref{VDM} can be proved by a direct calculation. By this lemma, solving
(\ref{Ben}) with respect to $H_{r}$ yields $n$ functions (Hamiltonians)
$H_{r}$ on $\mathcal{M}$
\begin{equation}
H_{r}=-\frac{1}{2}\sum_{i=1}^{n}\frac{\partial\rho_{r}}{\partial\lambda_{i}%
}\frac{f_{i}(\lambda_{i})\mu_{i}^{2}}{\Delta_{i}}+V_{r}(\lambda)\equiv\frac
{1}{2}\mu^{T}K_{r}G\mu+V_{r}(\lambda),\text{ \quad}r=1,\ldots,n \label{3.9}%
\end{equation}
called \emph{Benenti Hamiltonians}. Thus, for Benenti Hamiltonians the metric
tensor $G$ is given by
\[
G=\text{diag}\left(  \frac{f_{1}\left(  \lambda_{1}\right)  }{\Delta_{1}%
},\ldots,\frac{f_{n}\left(  \lambda_{n}\right)  }{\Delta_{n}}\right)
\]
while the Killing tensors $K_{r}$ are given by
\begin{equation}
K_{r}=-\text{diag}\left(  \frac{\partial\rho_{r}}{\partial\lambda_{1}}%
,\ldots,\frac{\partial\rho_{r}}{\partial\lambda_{n}}\right)  \quad
r=1,\ldots,n. \label{Kr}%
\end{equation}
From now and in what follows, we further assume that all $f_{i}$ are equal,
and likewise all $\varphi_{i}$:
\[
f_{i}:=f,\text{ \ }\varphi_{i}:=\varphi
\]
so that all the Hamiltonians (\ref{3.9}) are generated by the single
\emph{separation curve}:
\begin{equation}
\varphi(\lambda)+%
{\displaystyle\sum\limits_{j=1}^{n}}
\lambda^{n-j}H_{j}=\frac{1}{2}f(\lambda)\mu^{2} \label{Benc}%
\end{equation}
and are given explicitly by:
\begin{equation}
H_{r}=-\frac{1}{2}\sum_{i=1}^{n}\frac{\partial\rho_{r}}{\partial\lambda_{i}%
}\frac{f(\lambda_{i})\mu_{i}^{2}}{\Delta_{i}}+V_{r}(\lambda)\equiv\frac{1}%
{2}\mu^{T}K_{r}G\mu+V_{r}(\lambda),\text{ \quad}r=1,\ldots,n \label{4}%
\end{equation}
and thus the metric tensor $G$ is now given by
\[
G=\text{diag}\left(  \frac{f\left(  \lambda_{1}\right)  }{\Delta_{1}}%
,\ldots,\frac{f\left(  \lambda_{n}\right)  }{\Delta_{n}}\right)  .
\]
Of particular interest is the case $f(\lambda_{i})=\lambda_{i}^{m}$ with
$m\in\mathbb{Z}$. In such a case the metric tensor $G$ will be denoted by
$G_{m}$:
\[
G_{m}=\text{diag}\left(  \frac{\lambda_{1}^{m}}{\Delta_{1}},\ldots
,\frac{\lambda_{n}^{m}}{\Delta_{n}}\right)  \text{, }m\in\mathbb{Z}.
\]
Of course, if $f$ is a Laurent polynomial
\begin{equation}
\label{fpoly}f(\lambda)=\sum_{\alpha\in A}a_{\alpha}\lambda_{i}^{\alpha},
\end{equation}
where $A\subset\mathbb{Z}$ is a finite set, then
\begin{equation}
G=\sum_{\alpha\in A}a_{\alpha}G_{\alpha}. \label{GGm}%
\end{equation}
It can be shown that the metric $G_{m}$ is flat for $m\in\left\{
0,\ldots,n\right\}  $ and of constant curvature for $m=n+1$ (by linearity of
(\ref{Benc}) the same is true for $f$ being a polynomial in $\lambda$ of order
$m$). Moreover
\begin{equation}
G_{m}=L^{m}G_{0}\text{, \ }G_{0}=\text{diag}\left(  \frac{1}{\Delta_{1}%
},\ldots,\frac{1}{\Delta_{n}}\right)  , \label{GfromL}%
\end{equation}
where
\[
L=\text{diag}(\lambda_{1},\ldots,\lambda_{n})
\]
is a $(1,1)$-tensor called \emph{special conformal Killing tensor}%
~\cite{Crampin}. It can be shown~\cite{blasz2011} that all $K_{r}$ can be
calculated from the formula
\begin{equation}
K_{1}=I\text{, \quad}K_{r}=\sum\limits_{k=0}^{r-1}\rho_{k}L^{r-1-k}\text{,
\quad}r=2,\ldots,n. \label{KfromL}%
\end{equation}
In order to illustrate the form of separable potentials $V_{r}(\lambda)$ in
the Benenti case, we further assume that $\varphi$ is a Laurent sum of the
form
\begin{equation}
\varphi(\lambda)=\sum_{\alpha\in A}c_{\alpha}\lambda_{i}^{\alpha},
\label{Laurent}%
\end{equation}
where $A\subset\mathbb{Z}$ is a finite set and $c_{\alpha}$ are some real
constants. The Benenti separation relations (\ref{Ben}) become
\begin{equation}
\sum_{\alpha\in A}c_{\alpha}\lambda_{i}^{\alpha}+%
{\displaystyle\sum\limits_{j=1}^{n}}
\lambda_{i}^{n-j}H_{j}=\frac{1}{2}f(\lambda_{i})\mu_{i}^{2}\text{, \quad
}i=1,\ldots,n \label{Bens}%
\end{equation}
and due to their linearity we have
\[
V_{r}=\sum_{\alpha\in A}c_{\alpha}V_{r}^{(\alpha)},
\]
where $V_{r}^{(\alpha)}$ are so called \emph{basic separable potentials}. By
linearity of (\ref{Bens}), the potentials $V_{r}^{(\alpha)}$ satisfy the
relations
\[
\lambda_{i}^{\alpha}+\sum_{r=1}^{n}V_{r}^{(\alpha)}\lambda_{i}^{n-r}=0,\quad
i=1,\ldots,n
\]
and, again by Lemma \ref{VDM}, they are given by
\[
V_{r}^{(\alpha)}=\sum_{i=1}^{n}\frac{\partial\rho_{r}}{\partial\lambda_{i}%
}\frac{\lambda_{i}^{\alpha}}{\Delta_{i}},\text{ \quad}r=1,\ldots,n.
\]
The basic separable potentials $V_{r}^{(\alpha)}$ can be explicitly
constructed by the following formula \cite{blasz2011}:
\begin{equation}
V^{(\alpha)}=R^{\alpha}V^{(0)},\quad\ V^{(\alpha)}=(V_{1}^{(\alpha)}%
,\ldots,V_{n}^{(\alpha)})^{T}, \label{6}%
\end{equation}
where
\begin{equation}
R=\left(
\begin{array}
[c]{cccc}%
-\rho_{1} & \ 1 & \ 0 & \ 0\\
\vdots & \ 0 & \ \ddots & \ 0\\
\vdots & \ 0 & \ 0 & \ 1\\
-\rho_{n} & \ 0 & \ 0 & \ 0
\end{array}
\right)  \label{6a}%
\end{equation}
and $V^{(0)}=(0,\ldots,0,-1)^{T}$. The first $n$ basic potentials are trivial
\[
V_{k}^{(\alpha)}=-\delta_{k,n-\alpha}\text{, \quad}\alpha=0,\ldots,n-1.
\]
The first nontrivial positive potential is
\[
V^{(n)}=(\rho_{1},\ldots,\rho_{n})^{T}%
\]
and higher potentials are more complicated polynomials in $q_{i}$. The first
negative potential is
\[
V^{(-1)}=\left(  \frac{1}{\rho_{n}},\ldots,\frac{\rho_{n-1}}{\rho_{n}}\right)
^{T}%
\]
and the higher negative potentials are more complicated rational functions of
all $\rho_{i}$. Note also that the recursion formulas (\ref{6})-(\ref{6a}) are
not tensorial; they look the same in any coordinate system.

\section{Polynomial form of Benenti systems}

\label{sec.Polynomial form of Benenti systems} As we saw in the previous
section, even the relatively simple Benenti Hamiltonians are complicated
rational functions when expressed in the separation variables $(\lambda,\mu)$.
In this section we demonstrate two canonical maps that under certain
conditions transform Benenti Hamiltonians (\ref{4}) to a polynomial form.

\subsection{Benenti systems in Vi\`ete coordinates}

Suppose that we change the position coordinates on the base manifold
$\mathcal{Q}$ through the map
\begin{equation}
q_{i}=\rho_{i}(\lambda)\quad i=1,\ldots,n, \label{v1}%
\end{equation}
where, as in Lemma \ref{VDM}, $\rho_{i}(\lambda)=(-1)^{i}\sigma_{i}(\lambda)$.
This map induces the map (point transformation)\ on $T^{\ast}\mathcal{Q}$:
\begin{equation}
p=\left(  J_{V}^{-1}\right)  ^{T}\mu, \label{v2}%
\end{equation}
\label{v21} where $J_{V}$ is the Jacobian of the map (\ref{v1}):
\begin{equation}
\left(  J_{V}\right)  _{ij}=\frac{\partial\rho_{i}}{\partial\lambda_{j}}.
\end{equation}
Let us find an explicit form of (\ref{v2}). To do this we need the following lemma.

\begin{lemma}
\label{easy} Denote by $k_{i}$ the $i$-th column of an $n\times n$
nondegenerate matrix $A$:
\[
A=\left(  k_{1}|k_{2}|\ldots|k_{n}\right)
\]
and by $r_{j}$ the $j$-th row of its inverse
\[
A^{-1}=\left(
\begin{array}
[c]{c}%
r_{1}\\
r_{2}\\
\vdots\\
r_{n}%
\end{array}
\right)  .
\]
Then, if $\alpha_{i}\in\mathbb{R}$ for $i=1,\ldots,n$
\[
\left(  \alpha_{1}k_{1}|\alpha_{2}k_{2}|\ldots|\alpha_{n}k_{n}\right)
^{-1}=\left(
\begin{array}
[c]{c}%
r_{1}/\alpha_{1}\\
r_{2}/\alpha_{2}\\
\vdots\\
r_{n}/\alpha_{n}%
\end{array}
\right)  .
\]

\end{lemma}

This elementary lemma follows from the fact that $r_{i}k_{j}=\delta_{ij}$. An
analogous lemma is of course true if we consider rows of $A$ instead of its
columns. Combining lemmas \ref{VDM} and \ref{easy} we obtain that
\begin{equation}
\left(  J_{V}^{-1}\right)  _{ij}=-\frac{\lambda_{i}^{n-j}}{\Delta_{i}}
\label{v25}%
\end{equation}
and thus the map (\ref{v2}) can be written as
\begin{equation}
p_{i}=\ -\sum_{k=1}^{n}\frac{\lambda_{k}^{n-i}}{\Delta_{k}}\mu_{k},\quad
i=1,\ldots,n. \label{v3}%
\end{equation}
The coordinates $(q,p)$ defined by (\ref{v1}) and (\ref{v3}) are called
\emph{Vi\`{e}te coordinates}. To summarize, the map $(\lambda,\mu
)\rightarrow(q,p)$ from separation coordinates to Vi\`{e}te coordinates is
given by
\begin{equation}
q_{i}=\rho_{i}(\lambda)\text{, \ \ \thinspace}p_{i}=-\sum_{k=1}^{n}%
\frac{\lambda_{k}^{n-i}}{\Delta_{k}}\mu_{k},\ \ i=1,\ldots,n. \label{Viete}%
\end{equation}
Being a point transformation, the map (\ref{Viete}) is a canonical map which
means that Vi\`{e}te coordinates are Darboux (canonical) coordinates as well:
\[
\left\{  q_{i},q_{j}\right\}  =\left\{  p_{i},p_{j}\right\}  =0,\text{
\ }\left\{  q_{i},p_{j}\right\}  =\delta_{ij}.
\]
Let us now investigate the structure of Benenti Hamiltonians (\ref{4}) in
Vi\`{e}te coordinates $(q,p)$. The Hamiltonians (\ref{4}) are of course
written in tensor form so that in Vi\`{e}te coordinates
\begin{equation}
H_{r}(q,p)=\frac{1}{2}p^{T}K_{r}(q)G(q)p+V_{r}(q),\text{ \quad}r=1,\ldots,n
\label{BenViet}%
\end{equation}
where, by transformation laws for tensors,
\begin{equation}
K_{r}(q)=J_{V}K_{r}\left(  J_{V}\right)  ^{-1}\text{, \ }G(q)=J_{V}G\left(
J_{V}\right)  ^{T}. \label{transV}%
\end{equation}
The first formula in (\ref{transV}) yields, after some calculation
\begin{equation}
\left(  K_{r}(q)\right)  _{j}^{i}=\left\{
\begin{array}
[c]{l}%
q_{i-j+r-1},\quad i\leq j\text{ and }r\leq j\\
\\
-q_{i-j+r-1},\quad i>j\text{ and }r>j\\
\\
0\ \quad\text{otherwise}%
\end{array}
\right.  . \label{Krq}%
\end{equation}
Here and throughout the whole section we use the convention that $q_{0}=1$ and
$q_{k}=0$ for $k<0$ and for $k>n$. Thus, all the $K_{r}(q)$ are linear in
$q$-variables. Further, for the monomial case $f(\lambda_{i})=\lambda_{i}^{m}$
with $m\in\left\{  0,\ldots,n+1\right\}  $ we can obtain from the second
formula in (\ref{transV}) that
\begin{align}
G_{m}^{ij}  &  (q)\ =\left\{
\begin{array}
[c]{l}%
q_{i+j+m-n-1},\ \quad i,j=1,\ldots,n-m\\
\\
-q_{i+j+m-n-1},\ \quad i,j=n-m+1,\ldots,n\\
\\
0\ \quad\text{otherwise}%
\end{array}
\right.  \text{ \quad}m=0,\ldots,n\label{Gmq}\\
G_{m}^{ij}  &  (q)\ =q_{i}q_{j}-q_{i+j},\ \quad i,j=1,\ldots,n\text{, \ \quad
}m=n+1.\nonumber
\end{align}
The formulas (\ref{Krq}) and (\ref{Gmq}) can alternatively be obtained with
the help of the special conformal Killing tensor $L$ by using the formulas
(\ref{KfromL}) and (\ref{GfromL}), respectively, and the fact that the tensor
$L$ can be easily calculated in Vi\`{e}te coordinates through tensor
transformation law $L(q)=J_{V}L\left(  J_{V}\right)  ^{-1}$. We obtain
\[
L_{j}^{i}(q)=-\delta_{j}^{1}q_{i}+\delta_{j}^{i+1}%
\]
that is
\begin{equation}
L(q)=\left(
\begin{array}
[c]{cccc}%
-q_{1} & \ 1 & \ 0 & \ 0\\
\vdots & \ 0 & \ \ddots & \ 0\\
\vdots & \ 0 & \ 0 & \ 1\\
-q_{n} & \ 0 & \ 0 & \ 0
\end{array}
\right)  . \label{Linq}%
\end{equation}
Note therefore that $L$ happens to have the same form in $q$-coordinates as
the recursion matrix (\ref{6a}). This seems to be a pure coincidence without
any deeper meaning; we stress again that $R$ in (\ref{6a}) is not a tensor. In
any case, due to the fact that all the entries in $L$ are linear in $q_{i}$ we
see that all the entries in $G_{m}$ are linear in $q_{i}$ for $m=0,\ldots
,n+1$, quadratic in $q_{i}$ for $m=n+1$ and higher order polynomials for
higher $m$. Moreover, by (\ref{Krq}), all entries in $K_{r}(q)$ are linear in
$q_{i}$. Using all these facts and the formula (\ref{GGm}) we obtain the
following important corollary:

\begin{corollary}
\label{VC}If $f$ is a polynomial in (\ref{fpoly}), then the geodesic parts of
Benenti Hamiltonians (\ref{BenViet}) have a polynomial form. Moreover, if the
right hand side of (\ref{Laurent}) is a polynomial, then by the recursive
relations (\ref{6})-(\ref{6a}) also the potentials $V_{r}$ in the Benenti
Hamiltonians (\ref{4}) are in this case polynomials in $q_{i}$. Thus, in such
a case, the whole Hamiltonians $H_{r}(q,p)$ (and not just their geodesic
parts) are polynomials.
\end{corollary}

\begin{example}
\label{to Viete} Consider the case $n=2$, $f(\lambda)=1$ (i.e. a purely
monomial situation with $m=0$ in (\ref{Gmq}), so that $G=G_{0}$) and
$\varphi(\lambda)=\lambda^{3}$. Then the separation curve (\ref{Bens})
becomes
\[
\lambda^{3}+\lambda H_{1}+H_{2}=\frac{1}{2}\mu^{2}%
\]
and yields the Hamiltonians $H_{i}$ in the explicit form
\begin{align*}
H_{1} &  =\frac{1}{2(\lambda_{1}-\lambda_{2})}(\mu_{1}^{2}-\mu_{2}%
^{2})-(\lambda_{1}^{2}+\lambda_{1}\lambda_{2}+\lambda_{2}^{2})\\
H_{2} &  =\frac{1}{2(\lambda_{1}-\lambda_{2})}(\lambda_{1}\mu_{2}^{2}%
-\lambda_{2}\mu_{1}^{2})+\lambda_{1}\lambda_{2}(\lambda_{1}+\lambda_{2})
\end{align*}
so both Hamiltonians are rational functions of separation coordinates
$(\lambda,\mu)$. The above Hamiltonians have exactly the form (\ref{4}) with
the metric
\[
G=G_{0}=\text{diag}\left(  \frac{1}{\Delta_{1}},\frac{1}{\Delta_{2}}\right)
\text{,}%
\]
and with the Killing tensors (\ref{Kr}) given explicitly by:
\[
K_{1}=I\text{, \ }K_{2}=-\text{diag}(\lambda_{2},\lambda_{1}).
\]
The map (\ref{Viete}) to Vi\`{e}te coordinates has the explicit form:
\begin{align*}
q_{1} &  =-(\lambda_{1}+\lambda_{2})\text{, \ \ }q_{2}=\lambda_{1}\lambda
_{2},\\
p_{1} &  =\frac{1}{\lambda_{2}-\lambda_{1}}(\lambda_{1}\mu_{1}-\lambda_{2}%
\mu_{2})\text{, \quad}p_{2}=\frac{1}{\lambda_{2}-\lambda_{1}}(\mu_{1}-\mu
_{2}).
\end{align*}
An elementary calculations shows that $H_{i}$ in these variables attain the
form
\begin{align*}
H_{1}(q,p) &  =\frac{1}{2}q_{1}p_{2}^{2}+p_{1}p_{2}-q_{1}^{2}+q_{2}\\
H_{2}(q,p) &  =\frac{1}{2}p_{1}^{2}+q_{1}p_{1}p_{2}+\frac{1}{2}q_{1}^{2}%
p_{2}^{2}-\frac{1}{2}p_{2}^{2}q_{2}-q_{1}q_{2}%
\end{align*}
which is in agreement with (\ref{Gmq}) and (\ref{Krq}). Explicitly:
\[
G_{0}(q)=\left(
\begin{array}
[c]{cc}%
0 & 1\\
1 & q_{1}%
\end{array}
\right)  \text{, }K_{1}(q)=I\text{, \ }K_{2}(q)=\left(
\begin{array}
[c]{cc}%
0 & 1\\
-q_{2} & q_{1}%
\end{array}
\right)  .
\]
Thus, the Hamiltonians $H_{r}$ become polynomial in Vi\`{e}te coordinates
$(q,p)$.
\end{example}

\begin{example}
\label{3d}Consider the case $n=3$, $f(\lambda)=$ $\lambda$ (so that $m=1$ in
(\ref{Gmq}) and thus $G=G_{1}$) and $\varphi(\lambda)=\lambda^{5}$. Then the
separation curve (\ref{Bens}) becomes
\[
\lambda^{5}+\lambda^{2}H_{1}+\lambda H_{2}+H_{3}=\frac12\lambda\mu^{2}.
\]
Solving the corresponding separation coordinates yields the Benenti
Hamiltonians (\ref{4}) with the metric $G_{1}=LG_{0}$ with%
\[
L=\text{diag}\left(  \lambda_{1},\lambda_{2},\lambda_{3}\right)
\]
so that
\[
G_{1}=LG_{0}=\text{diag}\left(  \frac{\lambda_{1}}{\Delta_{1}},\frac
{\lambda_{2}}{\Delta_{2}},\frac{\lambda_{3}}{\Delta_{3}}\right)
\]
and with the Killing tensors (\ref{Kr}) given explicitly by
\begin{align*}
K_{1}  &  =I\text{, \quad}K_{2}=\text{diag}\left(  \lambda_{2}+\lambda
_{3},\lambda_{1}+\lambda_{3},\lambda_{1}+\lambda_{2}\right)  ,\\
K_{3}  &  =-\text{diag}\left(  \lambda_{2}\lambda_{3},\lambda_{1}\lambda
_{3},\lambda_{1}\lambda_{2}\right)  ,
\end{align*}
while the potentials $V_{r}=V_{r}^{(5)}$ have the form
\begin{align*}
V_{1}^{(5)}  &  =\lambda_{1}^{3}+\lambda_{2}^{3}+\lambda_{3}^{3}+\lambda
_{1}^{2}\lambda_{2}+\lambda_{1}^{2}\lambda_{3}+\lambda_{1}\lambda_{2}
^{2}+\lambda_{1}\lambda_{3}^{2}+\lambda_{2}^{2}\lambda_{3}+\lambda_{2}
\lambda_{3}^{2}+\lambda_{1}\lambda_{2}\lambda_{3},\\
V_{2}^{(5)}  &  =\lambda_{1}^{3}\lambda_{2}+\lambda_{1}^{3}\lambda_{3}
+\lambda_{1}^{2}\lambda_{2}^{2}+2\lambda_{1}^{2}\lambda_{2}\lambda_{3}
+\lambda_{1}^{2}\allowbreak\lambda_{3}^{2}+\lambda_{1}\lambda_{2}^{3}
+2\lambda_{1}\lambda_{2}^{2}\lambda_{3}\\
&  \phantom{=} +2\lambda_{1}\lambda_{2}\lambda_{3}^{2}+\lambda_{1}\lambda_{3}
^{3}+\allowbreak\lambda_{2}^{3}\lambda_{3}+\lambda_{2}^{2}\lambda_{3}
^{2}+\lambda_{2}\lambda_{3}^{3},\\
V_{3}^{(5)}  &  =\lambda_{1}\lambda_{2}\lambda_{3}\left(  \lambda_{1}
^{2}+\lambda_{2}^{2}+\lambda_{3}^{2}+\lambda_{1}\lambda_{2}+\lambda_{1}
\lambda_{3}+\lambda_{2}\lambda_{3}\right)  .
\end{align*}
The map (\ref{Viete}) to Vi\`ete coordinates has now the form
\[
q_{1}=-(\lambda_{1}+\lambda_{2}+\lambda_{3})\text{, \ \ }q_{2}=\lambda
_{1}\lambda_{2}+\lambda_{1}\lambda_{3}+\lambda_{2}\lambda_{3}\text{,
\ \ }q_{3}=-\lambda_{1}\lambda_{2}\lambda_{3}
\]
and
\[
\left(
\begin{array}
[c]{c}%
p_{1}\\
p_{2}\\
p_{3}%
\end{array}
\right)  =\left(  J_{V}^{-1}\right)  ^{T}\left(
\begin{array}
[c]{c}%
\mu_{1}\\
\mu_{2}\\
\mu_{3}%
\end{array}
\right)
\]
with $J_{V}$ and $J_{V}^{-1}$ given by (\ref{v21}) and (\ref{v25})
respectively. Explicitly
\[
J_{V}=\left(
\begin{array}
[c]{ccc}%
-1 & \ -1 & \ -1\\
\lambda_{2}+\lambda_{3} & \ \lambda_{1}+\lambda_{3} & \ \lambda_{1}
+\lambda_{2}\\
-\lambda_{2}\lambda_{3} & \ -\lambda_{1}\lambda_{3} & \ -\lambda_{1}
\lambda_{2}%
\end{array}
\right)
\]
and
\[
J_{V}^{-1}=-\left(
\begin{array}
[c]{ccc}%
\frac{\lambda_{1}^{2}}{\Delta_{1}} & \ \frac{\lambda_{1}}{\Delta_{1}} &
\ \frac1{\Delta_{1}}\\
\frac{\lambda_{2}^{2}}{\Delta_{2}} & \ \frac{\lambda_{2}}{\Delta_{2}} &
\ \frac1{\Delta_{2}}\\
\frac{\lambda_{2}^{2}}{\Delta_{3}} & \ \frac{\lambda_{2}}{\Delta_{3}} &
\ \frac1{\Delta_{3}}%
\end{array}
\right)  .
\]
An elementary calculation shows that $H_{i}$ in these variables attain the
form
\begin{align*}
H_{1}(q,p)  &  =\frac12q_{1}p_{2}^{2}+p_{1}p_{2}-\frac12q_{3}p_{3} ^{2}%
+q_{1}^{3}-2q_{1}q_{2}+q_{3},\\
H_{2}(q,p)  &  =\frac12p_{1}^{2}-\frac12q_{2}p_{2}^{2}-\frac12q_{1}q_{3}%
p_{3}^{2}+q_{1}p_{1}p_{2}-q_{3}p_{2}p_{3}+q_{1}^{2}q_{2} -q_{1}q_{3}-q_{2}%
^{2},\\
H_{3}(q,p)  &  =-\frac12q_{3}p_{2}^{2}-\frac12q_{2}q_{3}p_{3} ^{2}-q_{3}%
p_{1}p_{3}-q_{1}q_{3}p_{2}p_{3}+q_{1}^{2}q_{3}-q_{2}q_{3},
\end{align*}
which is in agreement with (\ref{Gmq}) and (\ref{Krq}). Explicitly:
\begin{align*}
G_{0}(q)  &  =\left(
\begin{array}
[c]{ccc}%
0 & 0 & 1\\
0 & 1 & q_{1}\\
1 & q_{1} & q_{2}%
\end{array}
\right)  \text{, }K_{1}(q)=I\text{, }K_{2}(q)=\left(
\begin{array}
[c]{ccc}%
0 & 1 & 0\\
-q_{2} & q_{1} & 1\\
-q_{3} & 0 & q_{1}%
\end{array}
\right)  ,\\
K_{3}(q)  &  =\left(
\begin{array}
[c]{ccc}%
0 & 0 & 1\\
-q_{3} & 0 & q_{1}\\
0 & -q_{3} & q_{2}%
\end{array}
\right)  ,
\end{align*}
while the tensor $L$ attains the form as in (\ref{Linq}):
\[
L(q)=\left(
\begin{array}
[c]{ccc}%
-q_{1} & \ 1 & \ 0\\
-q_{2} & \ 0 & \ 1\\
-q_{3} & \ 0 & \ 0
\end{array}
\right)  .
\]
Note again that the Hamiltonians $H_{r}$ become polynomial in Vi\`ete
coordinates $(q,p)$.
\end{example}

\subsection{Benenti systems in Newton coordinates}

The second method of turning Benenti Hamiltonian systems~(\ref{4}) into a
polynomial form is by using Newton coordinates. This method has been
discovered by V. M. Buchstaber and A. V. Mikhailov~\cite{buchstaber2018} only
quite recently. In this section we present our own proof of this result,
independent of the work \cite{buchstaber2018}. We also investigate in detail
the structure of Benenti Hamiltonians (\ref{4}) in Newton coordinates.

Consider the following map (consisting of a sequence of Newton polynomials) on
the base manifold $\mathcal{Q}$:
\begin{equation}
Q_{i}=\frac{1}{i}\sum\limits_{s=1}^{n}\lambda_{s}^{i}. \label{n1}%
\end{equation}
This map induces the map on $T^{\ast}\mathcal{Q}$:
\begin{equation}
P=\left(  J_{N}^{-1}\right)  ^{T}\mu, \label{n2}%
\end{equation}
where $P=(P_{1},\ldots,P_{n})^{T}$ and $J_{N}$ is the Jacobian of the map
(\ref{n1}),
\[
\left(  J_{N}\right)  _{ij}=\frac{\partial Q_{i}}{\partial\lambda_{j}}%
=\lambda_{j}^{i-1}.
\]
Thus, $J_{N}=V^{T}$, where $V$ is the Vandermonde matrix, but different from
$S$:
\begin{equation}
V=\left(
\begin{array}
[c]{cccc}%
1 & \lambda_{1} & \ldots & \lambda_{1}^{n-1}\\
\vdots & \vdots & \ddots & \vdots\\
1 & \lambda_{n} &  & \lambda_{n}^{n-1}%
\end{array}
\right)  . \label{W}%
\end{equation}
This also means that (\ref{n2}) leads to $P=V^{-1}\mu$.

\begin{lemma}
\label{Vlemma}In the above notation
\[
\left(  V^{-1}\right)  _{ij}=-\frac{1}{\Delta_{j}}\frac{\partial\rho_{n-i+1}%
}{\partial\lambda_{j}}.
\]

\end{lemma}

The reader should compare this lemma with Lemma \ref{VDM}. Thus, the map
(\ref{n1}) induces the following map on $T^{\ast}\mathcal{Q}$
\begin{equation}
Q_{i}=\frac{1}{i}\sum\limits_{s=1}^{n}\lambda_{s}^{i}\text{, \ }P_{i}%
=-\sum\limits_{j=1}^{n}\frac{1}{\Delta_{j}}\frac{\partial\rho_{n-i+1}%
}{\partial\lambda_{j}}\mu_{j}\text{, \ }i=1,\ldots,n\label{Newton}%
\end{equation}
and we call the coordinates $(Q,P)$ \emph{Newton coordinates} on $\mathcal{M}%
$. The reader should compare this map with the map (\ref{Viete}). Again, since
the map $(\lambda,\mu)\rightarrow(Q,P)$ is a point transformation map on
$T^{\ast}\mathcal{Q}$, the Newton coordinates $(Q,P)$ are Darboux (canonical)
coordinates, that is
\[
\left\{  Q_{i},Q_{j}\right\}  =\left\{  P_{i},P_{j}\right\}  =0,\text{
\ }\left\{  Q_{i},P_{j}\right\}  =\delta_{ij}.
\]
Let us now investigate the structure of Benenti Hamiltonians (\ref{4}) in
$(Q,P)$-coordinates. The Hamiltonians (\ref{4}) are written in tensor form and
thus
\begin{equation}
H_{r}(Q,P)=\frac{1}{2}P^{T}K_{r}(Q)G(Q)P+V_{r}(Q),\text{ \quad}r=1,\ldots
,n.\label{BenNewt}%
\end{equation}
In the monomial case, i.e., when $f(\lambda)=\lambda^{m}$ we have
\begin{equation}
H_{r}(Q,P)=\frac{1}{2}P^{T}K_{r}(Q)L^{m}(Q)G_{0}(Q)P+V_{r}(Q),\quad
r=1,\ldots,n.\label{BNmon}%
\end{equation}
Let us now investigate the structure of (\ref{BenNewt}) and in particular
(\ref{BNmon}), in Newton coordinates. Due to tensor transformation laws,
$L(Q),$ $K_{r}(Q)$ and $G(Q)$ are given by%
\begin{equation}
L(Q)=J_{N}L\left(  J_{N}\right)  ^{-1}\text{, \quad}K_{r}(Q)=J_{N}K_{r}\left(
J_{N}\right)  ^{-1}\label{p1}%
\end{equation}
and by
\begin{equation}
G(Q)=J_{N}G\left(  J_{N}\right)  ^{T}.\label{p2}%
\end{equation}
In order to express explicitly the right hand sides of (\ref{p1} ) and
(\ref{p2}) we need to invert the map $\lambda\rightarrow Q$ given by
(\ref{n1}), which is in general not algebraically invertible. Let us thus
consider the map $q\rightarrow Q$ between the Vi\`{e}te
coordinates~\eqref{Viete} and the Newton coordinates. In the recent
paper~\cite{BDM} this map is given by
\begin{equation}
Q_{r}=-\frac{1}{r}\sum_{k=1}^{r}V_{k}^{(n+r-k)}(q),\quad r=1,\dotsc
,n,\label{103}%
\end{equation}
where $V_{k}^{(\alpha)}(q)$ are the basic separable potentials as given by
(\ref{6})-(\ref{6a}). Below we present a theorem in which we considerably
extend the understanding of the above formula.

\begin{theorem}
\label{qQmap} The map $q\rightarrow Q$ as given by (\ref{103}) has the
following form:
\begin{equation}
\text{ }Q_{r}(q)=-q_{r}+\tau_{r}^{(r-1)}(q_{1},\ldots,q_{r-1})\text{,
\ \ }r=1,\ldots,n, \label{qQ}%
\end{equation}
where $\tau_{r}^{(\alpha)}$ denotes a polynomial of order $\alpha$ and where
$\tau_{1}^{(0)}=0$. The map $q\rightarrow Q\,$\ is algebraically invertible,
with the inverse map of the form
\begin{equation}
\text{ }q_{r}(Q)=-Q_{r}+\eta_{r}^{(r-1)}(Q_{1},\ldots,Q_{r-1})\text{,
\ \ }r=1,\ldots,n, \label{Qq}%
\end{equation}
where $\eta_{r}^{(\alpha)}$ denotes a polynomial of order $\alpha$ with
$\eta_{1}^{(0)}=0$. Moreover, neither $\tau_{r}^{(\alpha)}$ nor $\eta
_{r}^{(\alpha)}$ depends on $n$.
\end{theorem}

One proves this theorem by direct calculations, using the properties of basic
separable potentials $V_{k}^{(\alpha)}$. This theorem means that both the map
$q\rightarrow Q$ and its inverse $Q\rightarrow q$ are polynomial maps and
moreover that the transformation between the first $n$ variables, i.e. between
$q_{1},\ldots,q_{n}$ and $Q_{1},\ldots,Q_{n}$, does not change after
increasing $n$ to $n+1$. Expicitly, the first few expressions in both maps are%

\begin{align*}
Q_{1}  &  =-q_{1},\\
Q_{2}  &  =-q_{2}+\frac{1}{2}q_{1}^{2},\\
Q_{3}  &  =-q_{3}-\frac{1}{3}q_{1}^{3}+q_{2}q_{1},\\
Q_{4}  &  =-q_{4}+\frac{1}{4}q_{1}^{4}-q_{1}^{2}q_{2}+q_{3}q_{1}+\frac{1}%
{2}q_{2}^{2}\\
&  \vdots
\end{align*}
for the map $Q\rightarrow q$ and%
\begin{align*}
q_{1}  &  =-Q_{1},\\
q_{2}  &  =-Q_{2}+\frac{1}{2}Q_{1}^{2},\\
q_{3}  &  =-Q_{3}-\frac{1}{6}Q_{1}^{3}+Q_{2}Q_{1},\\
q_{4}  &  =-Q_{4}+\frac{1}{24}Q_{1}^{4}-\frac{1}{2}Q_{1}^{2}Q_{2}+Q_{3}%
Q_{1}+\frac{1}{2}Q_{2}^{2}\\
&  \vdots
\end{align*}
for the inverse map $q\rightarrow Q$. It is now possible to calculate the
tensor $L$ in the Newton coordinates $Q$. After some calculations we obtain:%

\begin{equation}
L(Q)=J_{N}L\left(  J_{N}\right)  ^{-1}=\left(
\begin{array}
[c]{ccccc}%
0 & 1 & 0 & \ldots & 0\\
0 & 0 & 1 & \ldots & 0\\
\vdots & \vdots & \vdots & \ddots & 1\\
-q_{n}(Q) & -q_{n-1}(Q) & -q_{n-2}(Q) & \ldots & -q_{1}(Q)
\end{array}
\right)  , \label{LQ}%
\end{equation}
or, equivalently
\[
L(Q)_{j}^{i}=-q_{n-j+1}(Q)\delta_{n}^{i}+\delta_{j-1}^{i}\text{, \quad
}i,j=1,\ldots,n,
\]
where the functions $q_{i}(Q)$ are given by (\ref{Qq}). Thus, the entries of
$L(Q)$ are polynomials, and the same is of course true for any positive power
$L^{m}(Q)$ of $L(Q)$.

Let us now calculate the Killing tensors $K_{r}$ in Newton coordinates $Q$. We
will do it by transforming $K_{r}(q)$, as given by (\ref{Krq}), to $Q$
variables, by the formula $K_{r}(Q)=J_{VN}K_{r}(q)\left(  J_{VN}\right)
^{-1}$, where $J_{VN}$ is the Jacobian transformation from Vi\`{e}te
coordinates to Newton coordinates. First we find that
\[
(J_{VN})_{i,j}=\displaystyle\sum\limits_{s=0}^{n}q_{s}(J_{VN})_{i-s,j}%
-q_{1}q_{i-s}+q_{i},\text{ \ }i,j=1,\ldots,n,\text{\ }%
\]
with $(J_{VN})_{1,j}=-1,\ (J_{VN})_{2,j}=q_{1},\ (J_{VN})_{3,j}=-q_{1}%
^{2}+q_{2}$ for any fixed $j$. This also yields that
\[
(J_{VN})_{i,j}^{-1}=-q_{i-j}\text{,\quad}i,j=1,\ldots,n.
\]
Note that this last result also means that the map (\ref{qQ}) can now be
extended to the whole manifold $\mathcal{M}=T^{\ast}\mathcal{Q}$ by completing
it with the map between the canonical momenta:
\begin{equation}
P_{i}=\left[  \left(  J_{VN\text{ }}\right)  ^{-1}\right]  _{ij}^{T}%
p_{j}=-\sum_{j=1}^{n}q_{j-i}p_{j},\ \ i=1,...,n. \label{pP}%
\end{equation}
After some calculations we obtain that
\begin{equation}
\left(  K_{r}(Q)\right)  _{j}^{i}=\left\{
\begin{array}
[c]{l}%
q_{i-j+r-1}(Q),\quad i-j\leq0\text{ and }r\leq n-i+1\\
\\
-q_{i-j+r-1}(Q),\quad i-j>0\text{ and }r>n-i+1\\
\\
0\quad\text{otherwise}%
\end{array}
\right.  , \label{KrQ}%
\end{equation}
cf. (\ref{Krq}). Thus, since all $q_{i}(Q)$ by (\ref{Qq}) are polynomials then
all the entries of $K_{r}(Q)$ are polynomials in $Q_{i}$ as well. Finally, let
us consider $G_{0}(Q)$, i.e. the metric $G_{0}$ in Newton coordinates, by
transforming $G_{0}(q)$, as given by (\ref{Gmq}), into $Q$ variables, by the
transformation formula $G_{0}(Q)=J_{VN}G_{0}(q)\left(  J_{VN}\right)  ^{T}$.

\begin{lemma}
\label{G0QL} The metric $G_{0}$ in Newton coordinates (\ref{n1}) attains the
form of a lower-triangular Hankel matrix given by the recursive formulas
\begin{equation}
G_{0}(Q)_{i,j}=\left\{
\begin{array}
[c]{c}%
-\sum\limits_{s=1}^{k}q_{s}(Q)\left(  G_{0}\right)  _{i-s,j}+q_{1}%
(Q)q_{i-1}(Q)-q_{i}(Q),\text{\quad}i\geq j\\
0,\text{ \quad}i<j
\end{array}
\right.  \text{ for }i,j=3,\ldots,n, \label{G0Q}%
\end{equation}
with $G_{0}(Q)_{1,j}=1,G_{0}(Q)_{2,j}=-q_{1}$, and $G_{0}(Q)_{3,j}=q_{1}%
^{2}-q_{2}$ for arbitrary fixed $j$.
\end{lemma}

As a consequence, the metric $G_{m}(Q)$ also attains the form of a
lower-triangular Hankel matrix. This can be verified using induction with
respect to $m$ in
\[
G_{m}(Q)_{i,j}=L(Q)_{j}^{i}G_{m-1}(Q)_{i,j}.
\]
Taking into account the formulas (\ref{LQ}), (\ref{KrQ}) and Lemma \ref{G0QL}
we obtain a corollary that is an analogue of Corollary~\ref{VC} for Newton coordinates.

\begin{corollary}
\label{NC} If $f$ is a polynomial in (\ref{fpoly}), then the geodesic parts of
Benenti Hamiltonians $H_{r}(Q,P)$ in (\ref{BenNewt}) have in Newton
coordinates (\ref{Newton}) a polynomial form. Moreover, if the right hand side
of (\ref{Laurent}) is a pure polynomial, then the potentials $V_{r}(Q)$ in the
Benenti Hamiltonians (\ref{BenNewt}) are in this case also polynomials. Thus,
in such a case, all the Hamiltonians $H_{r}(Q,P)$ (and not just their geodesic
parts) are polynomials.
\end{corollary}

Let us now present some examples.

\begin{example}
We proceed in the same setting as in Example \ref{to Viete}, i.e. we consider
the case $n=2$, $f(\lambda)=1$ (so that $m=0$) and $\varphi(\lambda
)=\lambda^{3}$, but in Newton coordinates. The map (\ref{qQ})-(\ref{pP}) reads
now
\begin{align*}
Q_{1}  &  =-q_{1},\text{ \ }Q_{2}=\frac{1}{2}q_{1}^{2}-q_{2},\\
P_{1}  &  =-p_{1}-q_{1}p_{2},\text{ \ }P_{2}=-p_{2}%
\end{align*}
and it transforms the Hamiltonians from Example \ref{Viete} to the form
\begin{align*}
H_{1}(Q,P)  &  =\frac{1}{2}P_{2}^{2}Q_{1}+P_{1}P_{2}-Q_{2}-\frac{1}{2}%
Q_{1}^{2},\\
H_{2}(Q,P)  &  =-\frac{1}{4}P_{2}^{2}Q_{1}^{2}+\frac{1}{2}P_{2}^{2}Q_{2}%
+\frac{1}{2}P_{1}^{2}+\frac{1}{2}Q_{1}^{3}-Q_{1}Q_{2},
\end{align*}
which is in agreement with (\ref{G0Q}) and (\ref{KrQ}). Explicitly:
\[
G_{0}(Q)=\left(
\begin{array}
[c]{cc}%
0 & \ 1\\
1 & \ Q_{1}%
\end{array}
\right)  \text{, }K_{1}(Q)=I\text{, }K_{2}(Q)=\left(
\begin{array}
[c]{cc}%
-Q_{1} & 1\\
Q_{2}-\frac{1}{2}Q_{1}^{2} & 0
\end{array}
\right)  .
\]
Moreover, $L$ becomes
\[
L(Q)=\left(
\begin{array}
[c]{cc}%
0 & 1\\
Q_{2}-\frac{1}{2}Q_{1}^{2} & Q_{1}%
\end{array}
\right)  .
\]

\end{example}

\begin{example}
We now consider Example \ref{3d} in Newton coordinates i.e. the case $n=3$,
$m=1$ and $\varphi(\lambda)=\lambda^{5}$. As $n=3$ the map (\ref{qQ}) is now
\begin{equation}
Q_{1}=-q_{1},\text{ \ }Q_{2}=\frac{1}{2}q_{1}^{2}-q_{2},\text{ \ }Q_{3}%
=-\frac{1}{3}q_{1}^{3}+q_{1}q_{2}-q_{3}, \label{qQ3}%
\end{equation}
and its inverse (\ref{Qq}) is
\[
q_{1}=-Q_{1},\text{ \ }q_{2}=\frac{1}{2}Q_{1}^{2}-Q_{2},\text{ \ }q_{3}%
=-\frac{1}{6}Q_{1}^{3}+Q_{1}Q_{2}-Q_{3}.
\]
The map (\ref{pP}) between momenta is
\begin{equation}
\left(
\begin{array}
[c]{c}%
P_{1}\\
P_{2}\\
P_{3}%
\end{array}
\right)  =\left(
\begin{array}
[c]{ccc}%
-1 & -q_{1} & -q_{2}\\
0 & -1 & -q_{1}\\
0 & 0 & -1
\end{array}
\allowbreak\right)  \left(
\begin{array}
[c]{c}%
p_{1}\\
p_{2}\\
p_{3}%
\end{array}
\right)  , \label{pP3}%
\end{equation}
with the inverse
\[
\left(
\begin{array}
[c]{c}%
p_{1}\\
p_{2}\\
p_{3}%
\end{array}
\right)  =\left(
\begin{array}
[c]{ccc}%
-1 & -Q_{1} & -\frac{1}{2}Q_{1}^{2}-Q_{2}\\
0 & -1 & -Q_{1}\\
0 & 0 & -1
\end{array}
\right)  \left(
\begin{array}
[c]{c}%
P_{1}\\
P_{2}\\
P_{3}%
\end{array}
\right)  .
\]
$\allowbreak$The map (\ref{qQ3})-(\ref{pP3}) transforms the Hamiltonians
$H_{r}(q,p)$ in Example \ref{3d} to the form
\begin{align*}
H_{1}(Q,P)  &  =\frac{1}{2}P^{T}\left(
\begin{array}
[c]{ccc}%
0 & 1 & Q_{1}\\
1 & Q_{1} & \frac{1}{2}Q_{1}^{2}+Q_{2}\\
Q_{1} & \frac{1}{2}Q_{1}^{2}+Q_{2} & Q_{3}+Q_{1}Q_{2}+\frac{1}{6}Q_{1}^{3}%
\end{array}
\right)  P+V_{1}^{(5)}(Q)\\
H_{2}(Q,P)  &  =\frac{1}{2}P^{T}\left(
\begin{array}
[c]{ccc}%
1 & 0 & Q_{2}-\frac{1}{2}Q_{1}^{2}\\
0 & Q_{2}-\frac{1}{2}Q_{1}^{2} & Q_{3}-\frac{1}{3}Q_{1}^{3}\\
Q_{2}-\frac{1}{2}Q_{1}^{2} & Q_{3}-\frac{1}{2}Q_{1}^{3} & -\frac{1}{12}%
Q_{1}^{4}-Q_{1}^{2}Q_{2}+Q_{3}Q_{1}+Q_{2}^{2}%
\end{array}
\right)  P+V_{2}^{(5)}(Q)\\
H_{3}(Q,P)  &  =\frac{1}{2}P^{T}\left(
\begin{array}
[c]{ccc}%
0 & 0 & \frac{1}{6}Q_{1}^{3}-Q_{2}Q_{1}+Q_{3}\\
0 & \frac{1}{6}Q_{1}^{3}-Q_{2}Q_{1}+Q_{3} & \frac{1}{6}Q_{1}^{4}-Q_{2}%
Q_{1}^{2}+Q_{3}Q_{1}\\
\frac{1}{6}Q_{1}^{3}-Q_{2}Q_{1}+Q_{3} & \frac{1}{6}Q_{1}^{4}-Q_{2}Q_{1}%
^{2}+Q_{3}Q_{1} &
\begin{array}
[c]{c}%
\frac{1}{12}Q_{1}^{5}-\frac{1}{3}Q_{1}^{3}Q_{2}+\frac{1}{2}Q_{3}Q_{1}^{2}\\
-Q_{1}Q_{2}^{2}+Q_{3}Q_{2}\allowbreak
\end{array}
\end{array}
\right)  P\\
&  \text{\ \ \ \ \thinspace}+V_{3}^{(5)}(Q),
\end{align*}

$\allowbreak$ where
\begin{align*}
V_{1}^{(5)}(Q)  &  =-\frac{1}{6}Q_{1}^{3}-Q_{1}Q_{2}-Q_{3},\\
V_{2}^{(5)}(Q)  &  =Q_{1}^{2}Q_{2}-Q_{1}Q_{3}-Q_{2}^{2}+\frac{1}{12}Q_{1}%
^{4},\\
V_{3}^{(5)}(Q)  &  =-\frac{1}{12}Q_{1}^{5}+\frac{1}{3}Q_{1}^{3}Q-\frac{1}%
{2}Q_{1}^{2}Q_{3}+Q_{1}Q_{2}^{2}-Q_{2}Q_{3},
\end{align*}
which is again in agreement with (\ref{G0Q}) and (\ref{KrQ}). Explicitly:
\begin{align*}
G_{0}(Q)  &  =\left(
\begin{array}
[c]{ccc}%
0 & \ 0 & \ 1\\
0 & \ 1 & \ Q_{1}\\
1 & \ Q_{1} & \ \frac{1}{2}Q_{1}^{2}+Q_{2}%
\end{array}
\right)  ,\quad K_{1}(Q)=I,\\
K_{2}(Q)  &  =\left(
\begin{array}
[c]{ccc}%
-Q_{1} & 1 & 0\\
0 & -Q_{1} & 1\\
\frac{1}{6}Q_{1}^{3}-Q_{1}Q_{2}+Q_{3} & Q_{2}-\frac{1}{2}Q_{1}^{2} & 0
\end{array}
\right)  ,\\
K_{3}(Q)  &  =\left(
\begin{array}
[c]{ccc}%
\frac{1}{2}Q_{1}^{2}-Q_{2} & -Q_{1} & 1\\
\frac{1}{6}Q_{1}^{3}-Q_{1}Q_{2}+Q_{3} & 0 & 0\\
0 & \frac{1}{6}Q_{1}^{3}-Q_{1}Q_{2}+Q_{3} & 0
\end{array}
\right)  ,
\end{align*}
and
\[
L(Q)=\left(
\begin{array}
[c]{ccc}%
0 & 1 & 0\\
0 & 0 & 1\\
\frac{1}{6}Q_{1}^{3}-Q_{1}Q_{2}+Q_{3} & Q_{2}-\frac{1}{2}Q_{1}^{2} & Q_{1}%
\end{array}
\right)  .
\]

\end{example}

\section{Conclusions}

\label{sec.Discussion and Conclusion} In this paper we have considered Benenti
Hamiltonian systems generated by a single separation curve. These systems turn
out to have a rational form when expressed in their separation coordinates.
Under certain additional conditions they can be cast into polynomial form
using either Vi\`{e}te coordinates (\ref{Viete}) or Newton coordinates
(\ref{Newton}), the last result due to Buchstaber and
Mikhailov~\cite{buchstaber2018}. We have presented a new version of Buchstaber
and Mikhailov results: not only have we proven their result in a more explicit
way but we also presented the explicit form of all the geometric objects,
associated with Benenti Hamiltonians, in Newton coordinates. This has been
done by constructing and analysing the map between the Vi\`{e}te and Newton coordinates.

A natural questions that arises is whether it is possible to extend our
construction to the case that $H_{i}$ are not generated by a single separation
curve but by the more general separation relations (\ref{Ben})\ i.e. with
different $f_{i}$ and $\varphi_{i}$. This will be a subject of another
research paper.

\subsection*{Acknowledgements}

The research of J.D. Maniraguha and C. Kurujyibwami was supported by
International Science Programme (ISP, Uppsala University) in collaboration
with Eastern Africa Universities Mathematics Programme (EAUMP). The research
of K. Marciniak was partially supported by the Swedish International
Development Cooperation Agency (Sida) through the Rwanda-Sweden bilateral
research cooperation.


\begin{thebibliography}{99}                                                                                               %


\bibitem {blaszak2006}B\l aszak M., Marciniak K., From St\"ackel systems to
integrable hierarchies of PDE's: Benenti class of separation relations,
\textit{J. Math. Phys.}, 2006, \textbf{47}, 3, p.032904.

\bibitem {blaszak2019}B\l aszak, M., \textit{Quantum versus Classical
Mechanics and Integrability Problems: towards a unification of approaches and
tools}, Springer, 2019.

\bibitem {blaszak2020}B\l aszak M., Marciniak K., St\"ackel transform of Lax
equations, \textit{Stud. Appl. Math.}, 2020.

\bibitem {BDM}B\l aszak M., Marciniak K., Domanski Z., Systematic construction
of non-autonomous Hamiltonian equations of Painlev\'e-type. I. Frobenius
integrability, 2020, arXiv:2001.02881.

\bibitem {blasz2011}B\l aszak M., Sergyeyev A., Generalized St\"ackel systems,
\textit{J. Math. Phys.}, 2011, \textbf{375}, 27, pp. 2617--2623.

\bibitem {BDSSB2013}B\l aszak M., Doma{\'{n}}ski Z., Sergyeyev A. and
Szablikowski BM., Integrable quantum St\"{a}ckel systems, \textit{Phys. Lett.
A}, 2013, \textbf{377}, 38, pp.2564-2572.

\bibitem {buchstaber2018}Buchstaber V. M., Mikhailov A. V., Polynomial
Hamiltonian integrable systems on symmetric powers of plane curves,
\textit{Russ. Math. Surv.}, Turpion Ltd, 2018, \textbf{73}, 6, pp. 1122--1124.



\bibitem {Crampin}Crampin M., W. Sarlet W., A class of nonconservative
Lagrangian systems on Riemannian manifolds, \textit{J. Math. Phys.}, 2001,
\textbf{42}, 9, pp. 4313--4326.

\bibitem {Stackel1891}St\"{a}ckel P., Die Integration der Hamilton-Jacobischen
Differentialgleichung mittelst Separation der Variablen, \textit{
Habilitationsschrift,} Halle, (1891), https://archiv.ub.uni-heidelberg.de/volltextserver/12758/.

\bibitem {sklyanin1995}Sklyanin E. K., Separation of variables-- new trends,
\textit{Progr. Theoret. Phys. Suppl.}, 1995, \textbf{118}, pp. 35--60.

\bibitem {waksjo2003}Waksj\"o C., Rauch-Wojciechowski S., How to find
separation coordinates for the Hamilton Jacobi equation: a criterion of
separability for natural Hamiltonian systems, \textit{J. Math. Phys. Anal.
Geom.}, 2003, \textbf{6}, pp.301--348.
\end{thebibliography}
\end{document}